\documentclass[%
reprint,
superscriptaddress,
 amsmath,amssymb,
 aps,
 pra,
]{revtex4-2}

\usepackage{graphicx}% Include figure files
\usepackage{dcolumn}% Align table columns on decimal point
\usepackage{bm}

\usepackage[utf8]{inputenc} 

\usepackage{multirow}
\usepackage[warn]{mathtext}
\usepackage{amsfonts}
\usepackage{amsmath}
\usepackage{amssymb}
\usepackage{braket}
\usepackage{bbold}
\usepackage{textcomp} 
\usepackage{indentfirst} 
\usepackage{amsmath} 
\usepackage{graphicx}
\usepackage{siunitx}
\usepackage{adjustbox}
\usepackage{tabularx}
\usepackage{ulem}

\DeclareGraphicsExtensions{.pdf,.png,.jpg}
\usepackage{hyperref}

\usepackage{algpseudocode}

\renewcommand{\bibname}{References}

\usepackage[nottoc]{tocbibind}
\usepackage{mathtools}

\usepackage{xcolor}
\hypersetup{
    colorlinks,
    linkcolor={red!50!black},
    citecolor={blue!50!black},
    urlcolor={blue!80!black}
}

\makeatletter

\newcommand{\us}{$\mu$s}
\newcommand{\fig}{Fig.~}

\makeatother

\begin{document}

    \title{
    Realization of a Quantum Error Detection Code with a Dynamically Reassigned Ancillary Qubit}

    \author{Alena~S.~Kazmina}
    \email[Corresponding author: ]{kazmina.as@phystech.edu}
    \affiliation{Russian Quantum Center, Skolkovo, Moscow 121205, Russia}
    \affiliation{National University of Science and Technology MISIS, Moscow 119049, Russia}
    \affiliation{Moscow Institute of Physics and Technology, Dolgoprudny 141700, Russia}

    \author{Artyom M.~Polyanskiy}
    \affiliation{Russian Quantum Center, Skolkovo, Moscow 121205, Russia}
    \affiliation{National University of Science and Technology MISIS, Moscow 119049, Russia}
    \affiliation{Moscow Institute of Physics and Technology, Dolgoprudny 141700, Russia}

    \author{Elena~Yu.~Egorova}
    \affiliation{Russian Quantum Center, Skolkovo, Moscow 121205, Russia}
    \affiliation{National University of Science and Technology MISIS, Moscow 119049, Russia}

    \author{Nikolay~N.~Abramov}
    \affiliation{National University of Science and Technology MISIS, Moscow 119049, Russia}

    \author{Daria~A.~Kalacheva}
    \affiliation{Skolkovo Institute of Science and Technology, Skolkovo Innovation Center, Moscow 121205, Russia}
    \affiliation{Moscow Institute of Physics and Technology, Dolgoprudny 141700, Russia}

    \author{Viktor~B.~Lubsanov}
     \affiliation{Moscow Institute of Physics and Technology, Dolgoprudny 141700, Russia}

    \author{Aleksey~N.~Bolgar}
     \affiliation{Moscow Institute of Physics and Technology, Dolgoprudny 141700, Russia}

    \author{Ilya~A.~Simakov}
    \affiliation{Russian Quantum Center, Skolkovo, Moscow 121205, Russia}
    \affiliation{National University of Science and Technology MISIS, Moscow 119049, Russia}
    \affiliation{Moscow Institute of Physics and Technology, Dolgoprudny 141700, Russia}

\date{\today}
 
    \begin{abstract}

    Quantum error correction (QEC) is essential for achieving fault-tolerant quantum computing. While superconducting qubits are among the most promising candidates for scalable QEC, their limited nearest-neighbor connectivity presents significant challenges for implementing a wide range of error correction codes. In this work, we experimentally demonstrate a quantum error detection scheme that employs a dynamically reassigned ancillary qubit on a chain of three linearly connected transmon qubits. We show that this scheme achieves performance comparable to conventional static-ancilla circuits. Additionally, the approach facilitates efficient quantum state preparation, which we demonstrate with tomography of arbitrary logical states. Our results provide a flexible method for implementing QEC codes under connectivity constraints and highlight a promising path toward scalable quantum architectures.
    \end{abstract}
	
\maketitle

 %%%% FIGURE 1 %%%%
\begin{figure*}
\includegraphics[width=0.99\linewidth]{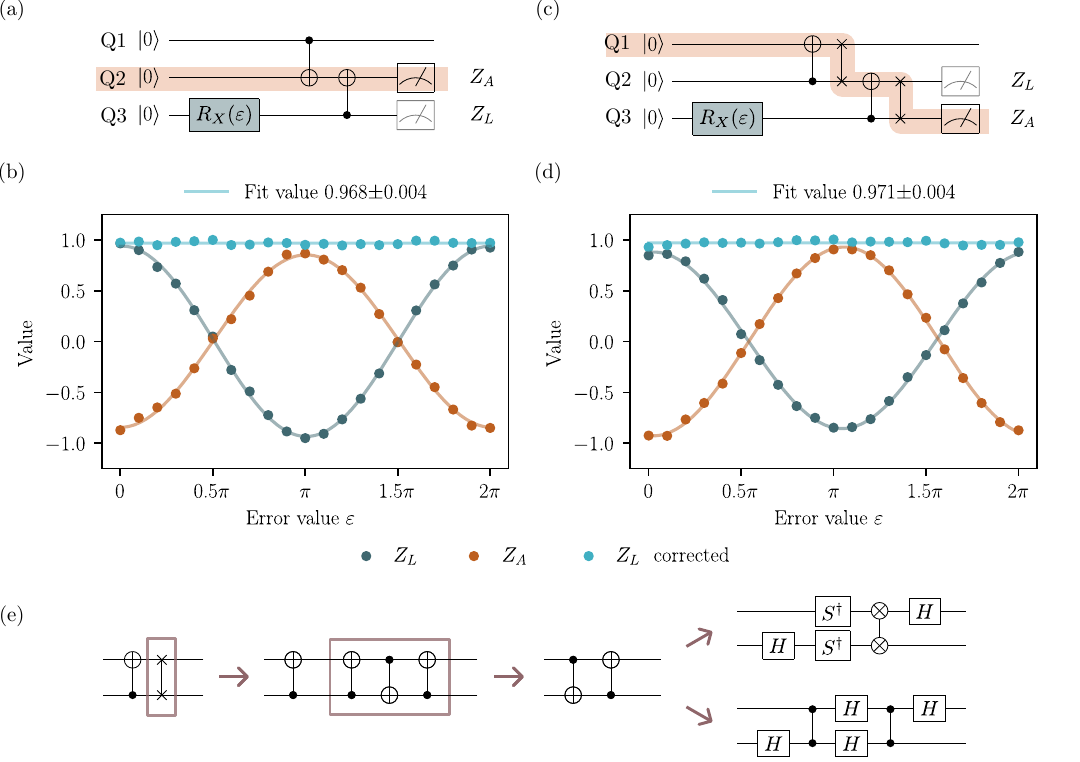}
\caption{
(a) The standard $X$-type error detection circuit and (b) the average measured ancilla outcome (orange), raw logical qubit observable (dark cyan) and error-corrected logical qubit observable (blue) as functions of an injection error $\varepsilon$.
(c) The gate sequence for a walking ancillary qubit scheme. The orange-highlighted area indicates the qubit that contains the ancillary qubit information throughout the circuit.
(d) Average auxiliary qubit outcome, raw logical qubit measurement outcome, and corrected logical qubit measurement outcome for the sliding ancillary qubit circuit.
(e) Possible implementations of the a CNOT gate followed by a SWAP gate with a single iSWAP gate or with two CZ gates.
}

\label{fig:error_detection_circuits_with_data}
\end{figure*}

%%%% FIGURE 2 %%%%
\begin{figure*}
\includegraphics[width=0.99\linewidth]{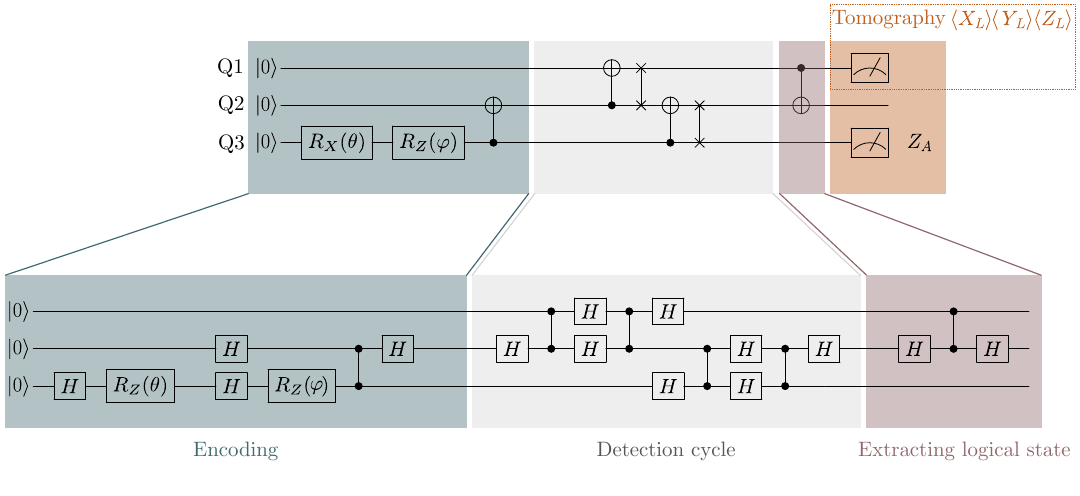}
\caption{
The logical state tomography circuit, incorporating additional transpiled frames for the walking ancillary qubit scheme. In this error detection protocol, the logical state is encoded between neighboring qubits using an entangling CNOT gate for various azimuthal $\varphi$ and polar angles $\theta$ (green-highlighted are). Following the error detection cycle, quantum state tomography is performed on the logical state by conducting measurements of the data qubit in the $X$, $Y$, and $Z$ Pauli bases. To enable the extraction of the logical state parameters, an additional CNOT gate is employed. The reconstructed logical state is represented as
$\rho = \frac{1}{2}(I+\braket{X_L} \sigma_x + \braket{Y_L} \sigma_y  + \braket{Z_L} \sigma_z )$, where $\sigma_i$  denote the Pauli matrices and $\braket{X_L}, \braket{Y_L}, \braket{Z_L}$ correspond to the expectation values of the logical Pauli operators. Furthermore, the syndrome outcomes obtained from ancillary qubit measurements are utilized in the post-selection process to facilitate subsequent error correction.}
\label{fig:logical_qubit_tomography_protocol_without_error}
\end{figure*}

%%%% FIGURE 3 %%%%
\begin{figure}
\includegraphics[width=0.99\linewidth]{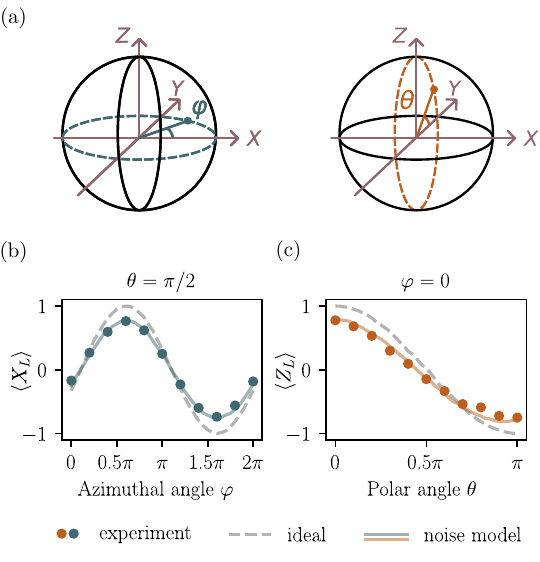}
\caption{Logical state tomography results. (a) Representation of azimuthal $\varphi$ and polar $\theta$ angles for the encoded logical state $\ket{\psi_{L}}$ in the Bloch sphere. (b) Extracted expectation value $\langle X_L \rangle$ of logical state for fixed polar angle $\theta = \pi / 2$ as function of azimuthal angle $\varphi$ (green dots). (c) The equivalent subfigure for expectation value $\langle Z_L \rangle$ as function of polar angle (orange dots). For both panels, the outcomes from the ideal unitary gate simulation are plotted with dashed gray lines. All experimental results are fitted using the noise model (green and orange solid lines) described in Appendix \ref{app:noise_model}.}
\label{fig:logical_qubit_tomography_V2}
\end{figure}

%%%% FIGURE 4 %%%%
\begin{figure}
\includegraphics[width=0.99\linewidth]{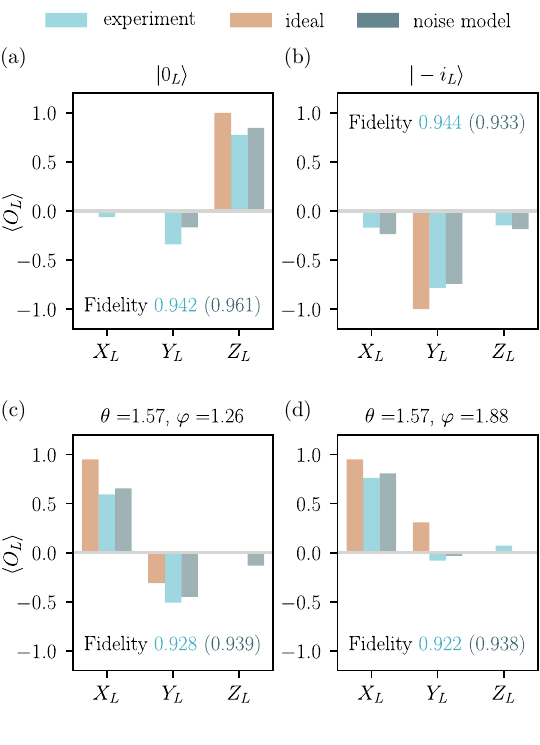}
\caption{The measured (blue), ideal (orange), and noise model simulated (green) expectation values of the Pauli logical operators for logical qubit expectation value are presented for the logical basis states $\ket{0_{L}}$ (a) and $\ket{-i_{L}}$ (b), and two additional states with parameters $\theta = 1.57, \varphi = 1.26$ (c) and $\theta = 1.57, \varphi = 1.88$ (d) close to $\ket{+_{L}}$. For each case, we show measured (blue), ideal (orange), and noise-model-simulated (green) values. State fidelities are calculated for both experimentally extracted values (blue) and simulated values (green, shown in brackets).}
\label{fig:pauli_logical_operators}
\end{figure}

%%%% FIGURE 5 %%%%
\begin{figure}
\includegraphics[width=0.99\linewidth]{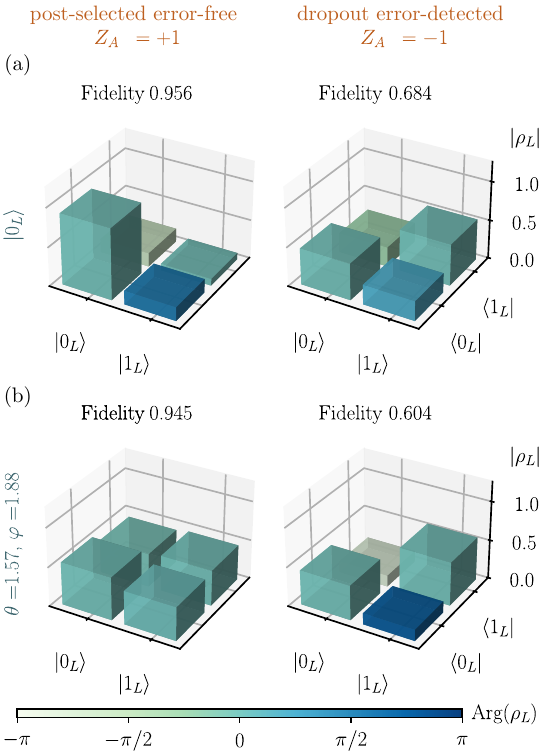}
\caption{Results of logical state tomography are shown for both error-free (left column) and error-detected (right column) cases, comparing: the prepared state $\ket{0_L}$ (a) and $\theta = 1.57, \varphi = 1.88$ (close to $\ket{+_L}$) (b). The calculated fidelity values for each reconstructed state highlight degraded outcomes in cases with $Z_A=-1$ syndrome measurements. This observed performance reduction correlates directly with the error-detection capability of the system.}
\label{fig:density_matrices}
\end{figure}

\section{Introduction} \label{sec:intro}

Quantum error correction (QEC) plays a crucial role in building scalable quantum processors capable of reliably performing computations with thousands or even millions of qubits \cite{PhysRevA.52.R2493, preskill1998fault, Calderbank_1996, DiVincenzoShor, SteaneQEC, kitaev1997quantum, Terhal_2015}. Stabilizer QEC codes achieve protection of quantum information using a redundant number of physical qubits \cite{Gottesman_1998}. A common approach to error correction relies on using physical qubits with two distinct roles: data qubits, which store the encoded logical quantum state, and ancillary qubits, which are periodically entangled with the data qubits to track errors through projective measurements of selected parity combinations of data qubits. The resulting measurement sequences are then analyzed to identify possible physical errors and correct the most likely logical ones.

To date, various QEC codes have been implemented across different hardware platforms. On superconducting qubits, there have been several demonstrations of repeated stabilizer measurements, memory experiments and simple operations on encoded qubits \cite{Krinner_2022, andersen2020repeated, chen2021exponentia, PhysRevLett.129.030501, DiCarlo2021logical, erhard2021entangling, besedin2025realizinglatticesurgerydistancethree, lacroix2024scaling}. For surface code \cite{kitaev2003fault, PhysRevA.86.032324}, the below-threshold regime has been achieved, where the error rate exponentially decreases with code distance \cite{google2023suppressing}. Moreover, a logical qubit has outperformed the best of physical qubits on which it was encoded \cite{acharya2024quantumerrorcorrectionsurface}. Alongside superconducting circuits, experiments have also been conducted on other promising physical systems, including trapped ions \cite{Nigg_2014, egan2021fault, hong2024entangling, paetznick2024demonstrationlogicalqubitsrepeated}, photons \cite{PRXQuantum.4.030340, vigliar2021error}, neutral atoms \cite{bluvstein2024logical, reichardt2024logicalcomputationdemonstratedneutral}, and silicon spin qubits \cite{Takeda_2022}. These diverse efforts demonstrate the applicability of quantum error correction across a range of hardware platforms and underline its importance for achieving scalable, fault-tolerant quantum computing.

Although superconducting devices are among the most promising for implementing quantum error correction, their restriction to nearest-neighbor connectivity between qubits complicates the realization of many error correction protocols. In contrast, platforms benefiting from all-to-all connectivity, such as trapped-ion systems, enable more hardware-efficient encodings of logical qubits, for example, low-density parity-check (LDPC) codes \cite{gottesman2014faulttolerantquantumcomputationconstant, PRXQuantum.2.040101}.

To address the connectivity constraints of superconducting processors for quantum error correction, one can adopt a dynamic code approach. A promising solution is originated on iSWAP operations \cite{schuch2003natural}, which simultaneously entangle qubits and exchange their computational states, enabling the dynamic reassignment of data and ancillary qubits within a quantum circuit. This technique was proposed for an efficient hardware implementation of the five-qubit code \cite{schuch2003natural, simakov2022, Antipov_2022, du2024fault} and its low-distance scaling \cite{Simakov_2025_low}. Experimentally, the idea of dynamic encoding has been demonstrated using a repeated surface code \cite{eickbusch2024demonstratingdynamicsurfacecodes}, showing that alternative non-fixed lattice structures can even outperform conventional layouts.

In this work, we experimentally demonstrate a quantum error detection code that employs a dynamically reassigned, or \textit{walking}, ancillary qubit, with its computational state transferred between physical qubits in a chain of three linearly connected transmons.
We provide a detailed description of how this walking ancilla scheme can be implemented using only native CZ two-qubit gates and show that its performance is comparable to that of a static-ancilla circuit. 
Moreover, we demonstrate that this method is particularly effective for quantum state preparation: since the data qubits start adjacent to each other, they can be directly entangled with high fidelity.

\color{black}
\section{Concept of a walking ancillary qubit}\label{sec:concept}
Conventional stabilizer QEC codes rely on the principle of syndrome measurement for redundantly encoded logical states using ancillary qubits. A basic circuit designed to detect a single-qubit $X$-type error is presented in \fig \ref{fig:error_detection_circuits_with_data}(a), where two CNOT gates are used to entangle an ancillary qubit Q2 with the two data qubits Q1 and Q3. The action of this circuit can be interpreted as a projective measurement of the $Z_{\mathrm{Q1}}Z_{\mathrm{Q3}}$ stabilizer with the ancillary qubit Q2.

An alternative error detection scheme (\fig \ref{fig:error_detection_circuits_with_data}(c)), proposed in this work, is specifically designed for devices with limited qubit connectivity and utilizes the concept of a walking ancillary qubit. During the detection cycle, the ancillary qubit quantum state is transferred from one physical qubit to another. Moreover, this approach allows the logical state to be encoded across neighboring data qubits, offering greater flexibility and compatibility with the architectural constraints of superconducting quantum processors.

We experimentally realize the proposed code using a three-qubit transmon-based device and compare the obtained results with that of the conventional error detection scheme. In both circuits presented in \fig \ref{fig:error_detection_circuits_with_data} (a) and (c), an $X$-type error is injected by applying a $R_X(\varepsilon)$ gate with a variable $\varepsilon$ angle, effectively acting as an error channel for the logical state $\ket{0_L}$, encoded into two physical qubits.
In the static-ancilla qubit scheme, two-qubit gates are performed between the ancillary qubit (Q2) and each of the two data qubits (Q1 and Q3). After the two-qubit gates, the stabilizer outcome $Z_\mathrm{A}$ can be read out in Q2. The logical qubit operator $Z_\mathrm{L}$ can be defined through either of the data qubits. Here we use the definition $Z_\mathrm{L} = Z_{Q3}$. For an error injected into Q3, we expect the outcome $Z_\mathrm{L}Z_\mathrm{A}$ to remain independent of the error angle $\varepsilon$. In the alternative dynamic error detection scheme, two additional SWAP gates transfer the ancillary qubit state from the physical qubit Q1 to Q3. Consequently, following the detection cycle, the stabilizer outcome $Z_A$ is measured at Q3, while the final logical qubit state remains encoded in the adjacent qubit pair Q1 and Q2. The syndrome extraction provides necessary information about the logical qubit state after the error injection, enabling further correction during post-measurement analysis based on the measured ancillary qubit outcome.
More details about the quantum device parameters, such as coherence times and gate performance, are given in the Appendix~\ref{app:device}. 

\fig \ref{fig:error_detection_circuits_with_data} presents the logical qubit observable, ancilla measurement results (orange), and reconstructed logical state, computed via the relation $Z_{L}^{{\text{corrected}}} = 1 - (p_{00} + p_{11})$ where $p_{ij}$, where $p_{ij}$ denotes the joint measurement probabilities of ancilla and data qubits. The comparison between (b) conventional and (d) walking error detection schemes shows excellent agreement in their performance.
Despite the increased circuit depth required for the walking error detection scheme compared to the conventional one, due to the transpilation of sequential CNOT and SWAP gates (\fig \ref{fig:error_detection_circuits_with_data}(e)), the efficiency of error correction in both cases remains comparable. This highlights the potential of walking circuits to offer robust error correction while maintaining a similar level of performance. However, it appears that the issue of larger circuit depth for the dynamic ancilla scheme can be addressed by using quantum processors with a native two-qubit iSWAP gate for superconducting processors \cite{ PhysRevLett.125.120504, PhysRevApplied.10.054062, PhysRevResearch.2.033447, PRXQuantum.5.020338}.

\section{Performance of a walking logical qubit} \label{sec:results}

As a characterization of the error detection scheme with the walking ancillary qubit, we additionally conducted a logical qubit state tomography experiment. The circuit of the experiment, along with its transpiled equivalent, is presented in \fig \ref{fig:logical_qubit_tomography_protocol_without_error}.
The scheme begins with the preparation of a logical qubit (green-highlighted area). A common method for encoding a logical state across multiple physical qubits is sequential entanglement, which enables to distribute quantum information \cite{NielsenChuang}. As a result, a logical qubit is defined within a two-dimensional subspace, and its state can be expressed in the following form
\begin{equation}
\ket{\psi_L} = \alpha \ket{00} + \beta \ket{11},
\label{eq:logical_state}
\end{equation}
where the coefficients $\alpha$ and $\beta$ can be parametrized with azimuthal $\theta$ and polar $\varphi$ angles, shown in \fig \ref{fig:logical_qubit_tomography_V2}(a): $\alpha=\cos \frac{\theta}{2},~\beta=e^{i \varphi} \sin \frac{\theta}{2}$.

After encoding the logical qubit, we assume that data qubits are subjected to error channels while performing gates throughout the sequence. These errors are subsequently identified via a detection cycle with a walking ancillary qubit (gray-colored area in \fig \ref{fig:logical_qubit_tomography_protocol_without_error}). The circuit also contains a supplementary CNOT gate that maps the logical qubit state into qubit Q1. To complete the protocol, the circuit executes standard quantum state tomography by measuring the logical qubit in the $X$, $Y$, and $Z$ bases. This measurements yield the expectation values $\langle X_L \rangle$, $\langle Y_L \rangle$, and $\langle Z_L \rangle$, which are used to reconstruct the logical state density matrix  $\rho = \frac{1}{2}(I+\braket{X_L} \sigma_x + \braket{Y_L} \sigma_y  + \braket{Z_L} \sigma_z )$, where $\sigma_i$ denote Pauli matrices. Such reconstruction enables full characterization of the logical state, providing the necessary information for quantifying its fidelity. Moreover, to detect the presence of bit-flip errors, we measure the syndrome $Z_A$ of the $ZZ$ stabilizer using an ancillary qubit.

The results of the logical state tomography are presented in \fig \ref{fig:logical_qubit_tomography_V2}(b, c), with green and orange dots. For better clarity, we provide data with fixed $\theta = \pi / 2$ and $\varphi=0$ for the azimuthal and polar angles, respectively.
We compare these values with outcomes from the noiseless simulation, plotted with gray dashed lines, and observe some differences both phase shift and amplitude for the logical state expectation values $\langle X_L \rangle$ and $\langle Z_L \rangle$ expectation.
To more thoroughly understand the discrepancies, we constructed a noise model reflecting the specific properties of the quantum hardware to interpret the experimental data. This
model accounts for both coherent errors, stemming from parasitic crosstalk and imperfections in two-qubit gates, and incoherent errors, arising from decoherence processes. The constructed model perfectly matches the experimental results, as illustrated in \fig \ref{fig:logical_qubit_tomography_V2}(b, c) with solid green and orange lines. Additional information about the noise model, its implementation and comparison with the experiment is provided in Appendix \ref{app:noise_model}.

To demonstrate application of the quantum error detection code with a walking ancillary qubit, we first compare the obtained density matrix $\rho$ with the expected ideal pure state and calculate its fidelity \cite{jozsa1994fidelity}. To further characterize performance, we compute the fidelity through noise-model simulations. Comparative analysis reveals good agreement between experimental and simulated results.
The measured logical qubit observables for two eigenstates of the logical $Z_L$ and $Y_L$ operators, $\ket{0_L}$ and $\ket{-i_L}$, as well as for two state vectors specified by parameters $\theta = 1.57, \varphi = 1.26$ and $\theta = 1.57, \varphi = 1.88$ are shown in \fig \ref{fig:pauli_logical_operators}.

Next, we evaluate the density matrix of the logical qubit conditioned on the measurement outcome of the stabilizer. In the case of no detected error ($Z_A=+1$), the density matrix is close to the ideal prepared state. Meanwhile, the density matrix postselected on runs where an error is detected ($Z_A=-1$) shows a significantly decreased state fidelity.
The classification results, including fidelity calculations, are presented in \fig \ref{fig:density_matrices}, demonstrate clear evidence of error-detection and post-selection capabilities. Additionally, we quantify the error-detection probability during circuit execution, which corresponds to the dropout fraction associated with the syndrome measurement outcome $Z_A=-1$, exhibiting an average value of 10 \%. 

\section{Conclusion and outlook} \label{sec:discussions}

In conclusion, we have experimentally realized a quantum error detection scheme that uses a walking ancillary qubit, whose computational state is dynamically transferred between physical qubits in a chain of three linearly connected transmons. Our implementation, based solely on native CZ two-qubit gates, achieves performance comparable to that of the conventional circuit. In addition, we demonstrated logical state tomography experiment, showing that this approach is beneficial for quantum state preparation: starting with adjacent data qubits allows for their high-fidelity entanglement and thus initial logical state preparation.

In a broader context, the realized error detection code stands out for its use of walking qubits, enabling dynamic reconfiguration of circuit topology. This feature is particularly valuable for quantum hardware with limited qubit connectivity in typical superconducting processors. Our approach aligns with recent advances in fault-tolerant circuits that exploit qubit movement and demonstrate circuit flexibility in QEC experiments \cite{eickbusch2024demonstratingdynamicsurfacecodes}. As a natural next step, this framework can be extended to support quantum error detection protocols involving multiple logical qubits \cite{Self_2024}, offering a scalable strategy to enhance the performance of quantum algorithms on NISQ devices and pave the way toward practical quantum computing.

\color{black}

\section*{Acknowledgements} 
We acknowledge Alexey Ustinov for fruitful discussions and comments on the manuscript. We sincerely thank Ilya Besedin for his invaluable contribution to the design of the superconducting device, as well as for his insightful discussions of the results and constructive feedback on the manuscript. This work was supported by Rosatom in the framework of the Roadmap for Quantum computing (Contract No. 868-1.3-15/15-2021 dated October 5). We acknowledge partial support of the Ministry of Science and Higher Education of the Russian Federation in the framework of the Program of Strategic Academic Leadership “Priority 2030” (Strategic Project Quantum Internet).
The transmon-based device was fabricated using the equipment of MIPT Shared Facilities Center.

% \clearpage
%%%% APPENDIX %%%%
\appendix

%%%% FIGURE 6 %%%%
\begin{figure*}
\includegraphics[width=0.99\linewidth]{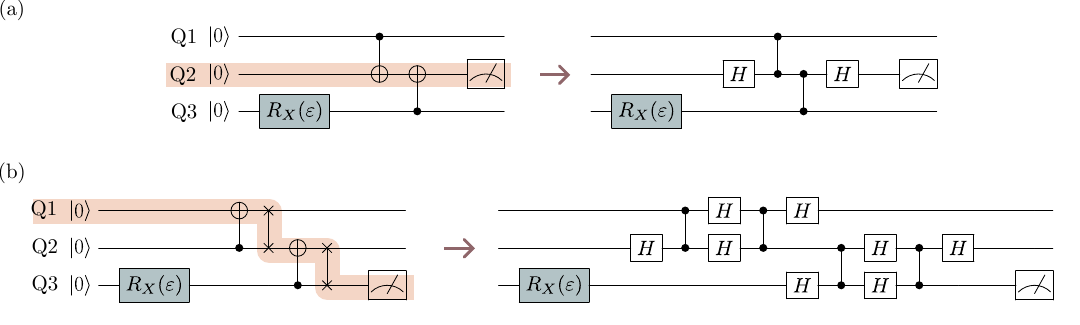}
\caption{Conventional (a) and sliding ancillary qubit (b) error detection circuits, along with their equivalents after the transpilation process for CZ and Hadamard gates, are shown. For clarity, the transpilation of Hadamard gates $H = 
R_Z\left(\frac{\pi}{2}\right)
R_X\left(\frac{\pi}{2}\right)
R_Z\left(\frac{\pi}{2}\right)$ is omitted, as well as the injected error gate $R_X(\varepsilon)=H R_Z(\varepsilon) H$}.
\label{fig:error_detection_circuits_transpilation}
\end{figure*}

%%%% FIGURE 7 %%%%
\begin{figure}\includegraphics[width=0.99\linewidth]{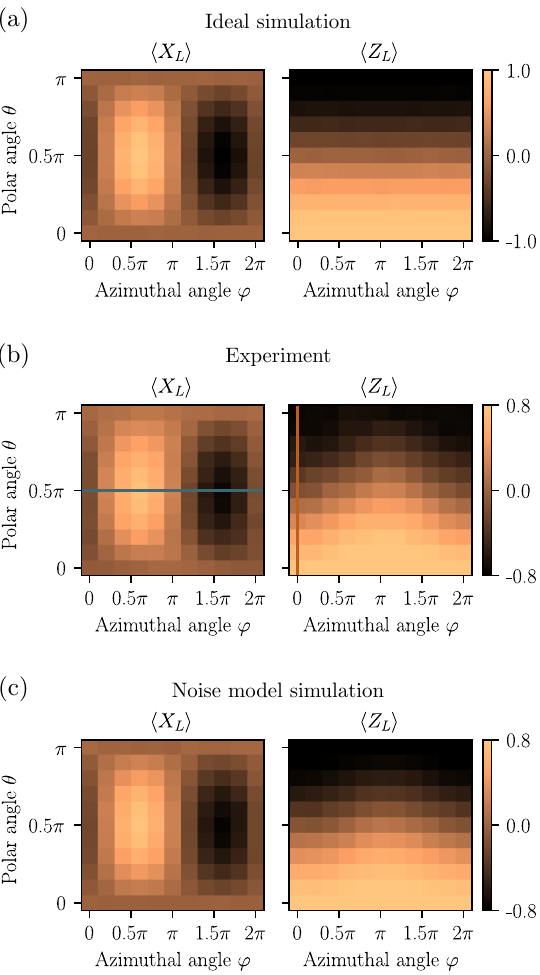}
\caption{(a) Unitary gate-based simulation of the logical qubit tomography experiment as a function of the input azimuthal $\varphi$ and polar $\theta$ angles, compared with experimental results (b) and noise model predictions (c). The green horizontal and vertical orange lines in subplot (b) indicate the cross-sections presented in \fig \ref{fig:logical_qubit_tomography_V2}(b) and (c) in the main text.}
\label{fig:3q_detection_tomography_theta_varphi}
\end{figure}

%%%% FIGURE 8 %%%%
\begin{figure*}
\includegraphics[width=0.99\linewidth]{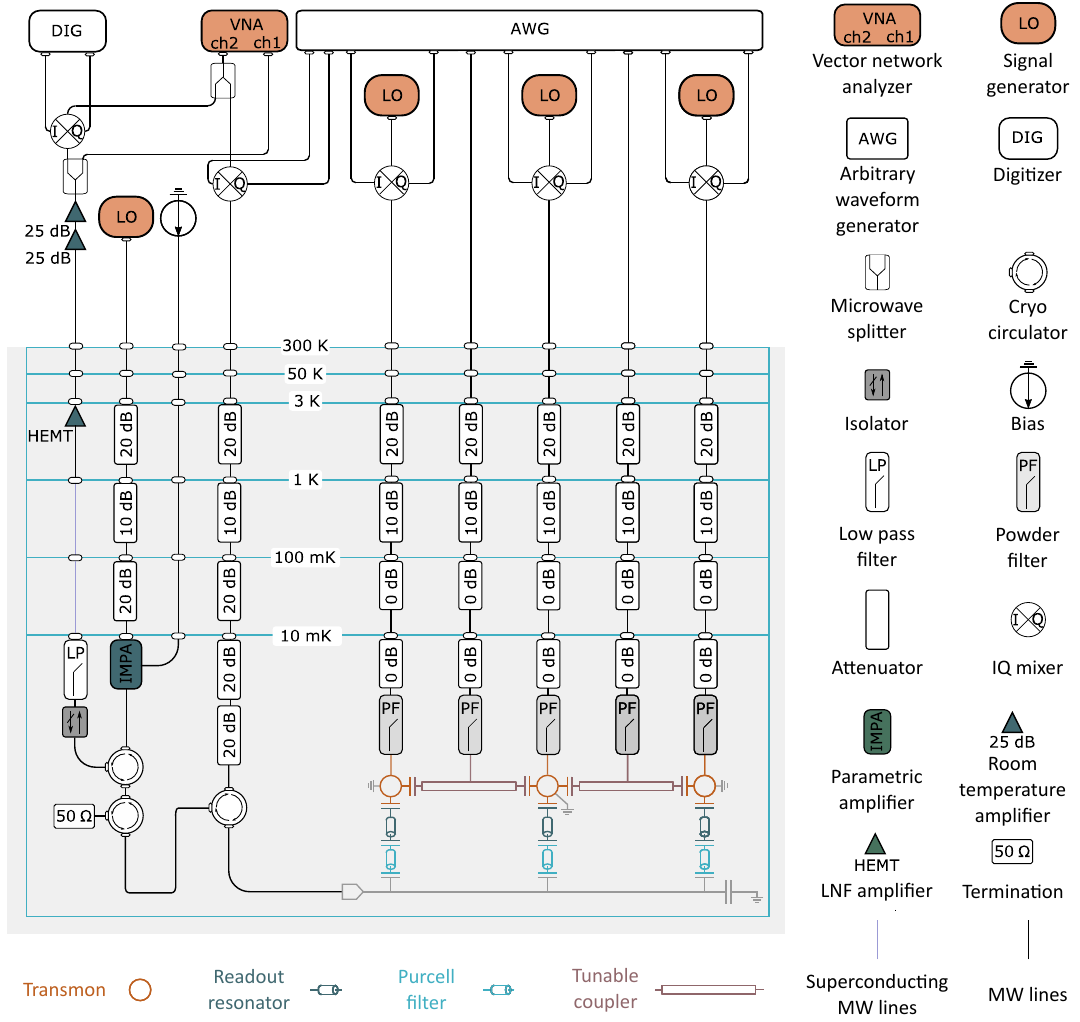}
\caption{The schematic of the experimental setup, with the cryogenic setup highlighted in the gray area. More information can be found in the main Appendix \ref{app:setup}.}
\label{fig:scheme_and_equip}
\end{figure*}

\section{Device} \label{app:device}

The experimental device contains three transmon-type qubits coupled to each other via a tunable coupling element, which allows the adjustment of the effective longitudinal $ZZ$ interaction and is used to perform controllable phase gates. More information about the unitary cell of the processor topology and gate implementation can be found in Ref. \cite{egorova2025threemodetunablecouplersuperconducting}. 
Each qubit in the device is dispersively coupled to its individual readout resonator and the main transmission line through a dedicated Purcell filter, which is utilized to enable faster multiplexed readout. The device was fabricated using standard techniques, including thin-film deposition, optical and e-beam lithography, as outlined in Ref. \cite{Kazmina_2024}.

All qubit parameters are determined through standard spectroscopy and time-domain measurements. To assess single-qubit gate fidelity, we utilize cross-entropy benchmarking technique (XEB) with single-qubit Clifford group reference sequences. Additionally, for benchmarking two-qubit CZ gates, we employed interleaved XEB. Such a technique allows to extract averaged single-qubit $1-p_1$ and two-qubit gate $1-p_2$ depolarization fidelities \cite{egorova2025threemodetunablecouplersuperconducting}  and calculate conventional ones using the expression \cite{Magesan_2012}
\begin{equation}
\label{eq:Fidelity}
    F = 1 - \frac{d-1}{d} p_i,
\end{equation}
where $d=2^n$ is a dimension of the relevant subspace of the Hilbert space for $n$ qubits and $i=1, 2$.

\begin{table*}[]
    \centering
    \caption{Qubit parameters, coherence properties and single-qubit performance.}
\begin{tabular}{lrrr}
\hline
\hline
Parameter &       Q1 &       Q2 &   Q3 \\
\hline
Qubit frequency $\omega_Q/2\pi$ (GHz)					        	        
& 6.264 & 6.360 & 6.172  \\

Relaxation time $T_1$ (\us)							                    
& 6.7 & 12.6 & 12.8  \\

Ramsey decay time $T_2$  (\us)				        	                
& 5.8 & 10.5 & 9.0 \\

Single qubit microwave gate $R_X (\pi / 2)$ duration (ns)		        		    
& 20 & 20 & 20 \\

Single-qubit gate fidelity (individual) $F_{\mathrm{1Q}}$ (\%)	    
& 99.71 & 99.89 & 99.85 \\

Readout frequency $\omega_\mathrm{RO}/2\pi$ (GHz)				        	    
& 7.400 & 7.240 & 7.104 \\

Readout duration $\tau_{\text{RO}}$ (ns)				        	    
& 700 & 700 & 700 \\

Readout fidelity (individual) $F_{\text{RO}}$ (\%)					        	    
& 88 & 83 & 93 \\

\hline
\hline
    \end{tabular}
    \label{tab:qubit_parameters}
\end{table*}

\begin{table*}[]
    \centering
    \caption{Two-qubit gates parameters.}
\begin{tabular}{lrrr}
\hline
\hline
Parameter &       Q1-Q2 &       Q1-Q2\\
\hline
CZ gate duration $\tau_{\text{CZ}}$ (ns)
& 55 & 48 \\
Single-qubit fidelity (simultaneous) $F_{\mathrm{1Q}}$ (\%) 
& 98.28 & 98.41\\
Two-qubit CZ gate fidelity $F_{\mathrm{CZ}}$ (\%) 
& $97.4 \pm 0.3$ & $97.9 \pm 0.3$\\

\hline
\hline
    \end{tabular}
    \label{tab:2q_gates_parameters}
\end{table*}

\section{Transpilation} \label{app:transpilation}

In the context of quantum processors, each device is typically designed to support a specific gate collection that can be physically implemented, commonly referred to as the native gate set. The quantum processor utilized in this experiment is constrained to the following native gates:  the single-qubit microwave-pulse rotations $R_X(\pi / 2)$,  the virtual $R_Z(\varphi)$ gates on arbitrary phase angle, and the two-qubit CZ operations. Therefore, transpilation process is based on a given input circuit to match the topology of a specific quantum device and required to adapt each algorithm for practical implementation.

The transpiled circuits for static and walking ancillary error detection cycles are depicted in \fig \ref{fig:error_detection_circuits_transpilation}(a, b), respectively. For clarity, the transpilation of Hadamard gates $H = 
R_Z\left(\frac{\pi}{2}\right)
R_X\left(\frac{\pi}{2}\right)
R_Z\left(\frac{\pi}{2}\right)$ has been omitted in this figure. The circuit depth of the walking ancillary code is twice larger than the conventional one. The increased depth contributes to the observed degradation of the results in the logical qubit experiment, as it makes the system more susceptible to decoherence effects resulting errors during the gates. Alternatively, the usage of a native iSWAP gate can provide lower circuit depth after the transpilation process overcoming this issue as discussed in the main text.

\section{Simulation} \label{app:noise_model}

To validate the results obtained from the logical qubit tomography, we simulate the transpired gate sequence presented in \fig \ref{fig:logical_qubit_tomography_protocol_without_error} using a circuit-level noise model.  The noise model includes two error mechanisms: coherent errors, arising from parasitic interactions between physical qubits, and incoherent errors, which can be attributed to decoherence processes.

To characterize the coherent errors in our model, we first establish an appropriate error model by analyzing the reduced two-level system Hamiltonian that describes the two-qubit interaction in the rotating frame, given by
\begin{equation} \label{eq:2Q_reduced_hamiltonian}
\begin{split}
    \frac{\mathcal{H}}{\hbar} = \frac{1}{4} \zeta (\Phi )\sigma_{z}^{(1)} \sigma_{z}^{(2)} + \frac{1}{2} J (\Phi) \Bigl( & e^{i \delta \omega t} \sigma_{x}^{(1)} \sigma_{x}^{(2)} + \\ & +
     e^{-i \delta \omega t} \sigma_{y}^{(1)} \sigma_{y}^{(2)} \Bigl) ,
\end{split}
\end{equation}
where $\delta \omega = \omega_{q1} - \omega_{q2}$ represents qubit detuning, $\zeta (\Phi)$ denotes the longitudinal $ZZ$ coupling between qubits, $J (\Phi)$ represents the transverse $XX$ coupling, which are both depends on external magnetic flux in a coupler $\Phi$,  and $\sigma_{x}^{(i)}$, $\sigma_{y}^{(i)}$ and $\sigma_{z}^{(i)}$ are Pauli operators.

The reduced Hamiltonian indicates the presence of two potential sources of crosstalk errors, which can be utilized in our noise model. In the first instance, it can be observed from (\ref{eq:2Q_reduced_hamiltonian}) that during single-qubit gates or idle periods, when the coupler is detuned far away from the physical qubits, non-zero parasitic longitudinal crosstalk may accumulate phase, leading to a coherent error. We model it by the unitary matrix of control phase gate

\begin{equation}
    \label{eq:cphase_matrix}
    \text{CPHASE}(\delta \varphi) =
    \begin{pmatrix}
    1 & 0 & 0 & 0 \\
    0 & 1 & 0 & 0 \\
    0 & 0 & 1 & 0 \\
    0 & 0 & 0 & e^{i\delta \varphi}
    \end{pmatrix}
\end{equation}
applied after all single-qubit $R_X(\pi / 2)$ gates.

In addition, during the execution of CZ gates, unwanted transverse interaction can induce coherent rotations between the $\ket{01}$ and $\ket{10}$ states. The unitary operator representing this error in the near-resonance regime takes the form of an $XY$ gate and is characterized by the matrix

\begin{equation}
    \label{eq:xx_plus_yy_matrix}
    R_{XX+YY}(\theta) = 
    \begin{pmatrix}
    1 & 0 & 0 & 0 \\
    0 & \cos \frac{\theta}{2} & -i \sin \frac{\theta} {2} & 0 \\
    0 & -i \sin \frac{\theta} {2} & \cos \frac{\theta}{2} & 0 \\
    0 & 0 & 0 & 1
    \end{pmatrix}.
\end{equation}
Here, the angle  $\theta$  is proportional to interaction strength $J$ in the Hamiltonian \ref{eq:2Q_reduced_hamiltonian} and two-qubit-gate duration. Since the $R_{XX+YY}(\theta)$ gate commutes with our native CZ gate, we incorporate this coherent error term following each CZ gate in the pulse sequence.

The second type of errors in the model arises purely from decoherence processes and is represented by a depolarization channel with parameter $p$. This error channel acting on a density matrix $\rho$ is defined as
\begin{equation}
    E(\rho) = p \rho + (1-p) \frac{I}{d},
\end{equation}
where $d=2^n$ is a dimension of the relevant computational subspace. Accordingly, we incorporate this error channel along with two distinct depolarization error probabilities: $p_1$ for all non-zero duration single-qubit gates and $p_2$ for two-qubit gates.

Additionally, it should be noted that we did not consider readout errors, as all our measurement results are corrected using a SPAM (state preparation and measurement) matrix.

We estimate the free parameters in the model $p_1$, $p_2$, $\theta$, and $\varphi$ by fitting the logical qubit expectation values $\braket{X_L}$ and $\braket{Z_L}$ to the described error model, employing a dual annealing optimization process. For model simplicity, we assume that all error values are identical for equivalent gates, with the depolarization parameters for both single-qubit and two-qubit gates being the same.
All extracted parameters are presented in Table \ref{tab:noise_model_params}. The noise model predictions for logical qubit expectation values, shown in Fig. \ref{fig:3q_detection_tomography_theta_varphi}(c), demonstrate good agreement with the corresponding experimental results presented in the panel (b).

Additionally, we compare the fidelity of the single-qubit and two-qubit gates obtained from this model to gate fidelities obtain from randomized benchmarking. To quantify the contribution of coherent errors, we employ the average fidelity metric for a d-dimensional quantum system, defined as
\begin{equation}
\label{eq:Fidelity_unitary}
    F = \frac{1}{d(d + 1)} 
    \left( \text{Tr} \left( U^\dagger U \right)  + |\text{Tr}(U_\mathrm{t} ^\dagger U)| ^ 2 \right),
\end{equation}
where $U$ represents the actual implemented unitary operation and $U_\mathrm{t}$ denotes the target unitary. This equation yields error expressions for different interaction types: the infidelity due to the static ZZ interaction with the accumulated phase $\delta \varphi$ on single qubit gates

\begin{equation}
\label{eq:error_ZZ_interaction}
    1 - F = \frac{3}{10} (1- \cos \delta \varphi),
\end{equation}
and the exchange interaction contribution to CZ gate infidelity with parameter $\theta$

\begin{equation}
\label{eq:error_exchange_interaction}
    1 - F = \frac{1}{20} (12 - 8\cos\frac{\theta}{2} - 4\cos ^ 2 \frac{\theta}{2}).
\end{equation}

For single-qubit gates, the noise model predicts contributions to the overall error: a decoherence-induced error due to depolarization $p_1$, quantified as $1.78 \cdot 10  ^ {-2}$, and an error arising from the static ZZ interaction, evaluated to be $10 ^ {-4}$ according to the equation (\ref{eq:error_ZZ_interaction}). The depolarization error can be translated into a corresponding gate fidelity using the expression (\ref{eq:Fidelity}), yielding a fidelity value of 98.6\%. This fidelity is in good agreement with the experimentally obtained fidelity derived from simultaneous randomized benchmarking, as reported in Table \ref{tab:2q_gates_parameters}.

A similar analysis can be performed for two-qubit gates. According to the noise model, this error arises from two primary contributions: depolarization processes characterized by the parameter $p_2=1.78 \cdot 10  ^ {-2}$, and an additional error of $1.17 \cdot 10 ^ {-2}$ due to the presence of exchange interaction, estimated by the equation (\ref{eq:error_exchange_interaction}). The combined effect of these errors corresponds to a gate fidelity of 97.4\%, calculated using equation (\ref{eq:Fidelity}). In this case, the total error predicted by the noise model is in close agreement with the experimentally measured fidelity of the two-qubit CZ gate, as presented in Table \ref{tab:2q_gates_parameters}.

\begin{table}[]
    \centering
    \caption{Noise model parameters extracted from the fitter process.}
\begin{tabularx}{\columnwidth} { 
     >{\raggedright\arraybackslash}X 
     >{\centering\arraybackslash}X 
     >{\centering\arraybackslash}X 
     >{\centering\arraybackslash}X
     >{\centering\arraybackslash}X}
\hline
\hline
Parameter &       $\delta \varphi$ &       $\theta$ & $p_1$ & $p_2$ \\
\hline
Value &       -0.027 &       0.37 & $0.0178$ & $0.0178$\\
\hline
\hline
\end{tabularx}
    \label{tab:noise_model_params}
\end{table}

\section{Experimental setup} \label{app:setup}
The experiment was performed in a dilution refrigerator operating at a base temperature of around 10 mK. The overall experimental configuration is organized into two primary sections: the room-temperature system and the cryogenic system, each comprising various associated components, as shown in \fig \ref{fig:scheme_and_equip}.

The transmon-based processor is mounted to the bottom plate of the dilution refrigerator and packaged in a custom-designed sample holder specifically intended for this setup. Thermal stabilization of the device is achieved through a series of cryogenic attenuators, which effectively mitigate unwanted heat flow from the room-temperature components by dissipation processes. In addition, incoming signals are filtered using custom RF powder filters. The qubits, represented as orange circles, are controlled through a common IQ up-conversion heterodyne scheme with a local oscillator (LO), supported by two channels of an arbitrary waveform generator (AWG) for pulse-shaping control. The tunable coupling element, indicated in grayed-pink, is manipulated by fast flux pulses originating from the same control device.

State discrimination is performed by applying a multiplexed rectangular pulse to the main readout line. The reflected pulse is processed through a three-stage amplification sequence, beginning with an impedance matching amplifier (IMPA), followed by a high-electron mobility transistor (HEMT), and concluding with room-temperature amplification. The resulting signal is then downconverted and processed using the appropriate integration function by an FPGA-based digitizer system (DIG). 

% \bibliographystyle{unsrtnat}
% \bibliography{library}
\clearpage
\renewcommand{\bibname}{Reference}
\normalem{}
\bibliography{library}

\end{document}